\documentstyle[prl,aps,epsf,times,float,graphicx]{revtex} 
\DeclareGraphicsExtensions{.eps,.ps} 
\begin{document} 
\draft 
\twocolumn[\hsize\textwidth\columnwidth\hsize\csname@twocolumnfalse\endcsname 
\title{The electronic specific heat in the pairing pseudogap regime} 
\author{C. P. Moca$^{1,2,3}$ and Boldizs\'ar Jank\'o$^{1,2}$} 
\address{
$^1$ Materials Science Division, Argonne National Laboratory, Argonne, Illinois 60439, USA\\
$^2$ Department of Physics, Notre Dame University, Notre Dame Indiana, 46556,USA \\
$^3$ Department of Physics, University of Oradea, Oradea, 3700, Romania 
}  
\date{\today} \maketitle 
%---------------------------------------------------------------- 
\begin{abstract} 
When pairing correlations in a quasi two dimensional electron system induce 
a pseudogap in the single particle density of states, the 
specific heat must also contain a sizeable pair contribution. The theoretically
calculated  specific heat for such a system is compared to the 
experimental results of Loram and his collaborators for underdoped $% 
YBa_2Cu_3O_{6+x}$ and $La_{2-x}Sr_{x}CuO_4$ samples. The size and doping 
dependence of the extracted pseudogap energy scale for both materials is 
comparable to the values obtained from a variety of other experiments. 
\end{abstract}   
] 
\makeatletter 
\global\@specialpagefalse 
\def\@oddhead{REV\TeX{} 3.0\hfill } 
\let\@evenhead\@oddhead 
\makeatother

\emph{Introduction. }The gradual, but substantial loss of low energy single 
particle states in the normal state of underdoped cuprate samples is by now 
documented by a large number of experiments \cite{timusk}. The onset of the
pseudogap regime is captured most clearly by spectroscopic probes that couple 
predominantly to the single particle excitations, such as tunneling \cite{tunneling} 
and angle-resolved photoemission spectroscopy $(ARPES)$ \cite{arpes}. All
theoretical  expectations, however, point towards some sort of many-particle
phenomenon  behind the pseudogap effect, and imply observable changes in 
collective electronic properties \cite{houston}. Traditional probes of collective 
degrees of freedom, such as inelastic neutron scattering \cite{neutron}, Raman 
scattering \cite{raman}, optical conductivity \cite{timusk}, were used extensively on 
pseudogapped samples, while new experiments, such as higher order tunneling 
spectroscopy \cite{tunn2}, were also proposed. Despite such remarkable
accumulation of high quality experimental data a consensus has yet to emerge
regarding the origin of the collective effect causing the pseudogap.
     
One of the earliest experimental indications of normal state
anomalies in underdoped cuprates came not from spectroscopic but
from thermodynamic measurements. Loram and his collaborators found
\cite{loram} that the coefficient of the electronic heat capacity
$\gamma _{\mathrm{el}}(T) = C_{\mathrm{el}}(T)/T$ of underdoped samples is
no longer constant in  temperature, as it is for the optimal and overdoped crystals,
and as one would  expect from a normal Fermi liquid. Instead,
$\gamma_{\mathrm{el}} (T)$ shows  a broad maximum at a crossover temperature
$T^*$ which is doping dependent  and typically much higher than the
superconducting transition temperature $ T_c$. As $T$ is lowered below $T^*$, 
$\gamma_{\mathrm{el}}(T)$ decreases significantly before the temperature
reaches $T_c$ . Despite the fact, that these specific
heat measurements have
been available for some time,  there is no systematic study of this data
within any theoretical framework of a proposed  pseudogap scenario. 
The purpose of this paper is to provide such an analysis within a
pairing fluctuation framework.

There are several
reasons why the results of these  thermodynamic measurements deserve a
detailed theoretical analysis. First,  there is ample experimental data on the
most extensively studied classes  of cuprate superconductors,
$YBa_2Cu_3O_{6+x}$, $La_{2-x}Sr_xCuO_4$ \cite{loram}, and recently
$Bi_2Sr_2CaCu_2O_{8+x}$ \cite{loram-biscco}, for a remarkably wide range of
doping.  This makes it possible to single  out and to concentrate the
theoretical study on the universal features mentioned  above. Second,
these thermodynamic measurements  couple to  all  thermally
excitable modes, irrespective of  their single particle, or collective
character. In  tunneling and photoemission the higher order
excitations, such as pair  excitations, may be detected only through their
convoluted influence on the  single particle spectrum. Thus, in contrast to
any theoretical analysis of  spectroscopic data, where one  seems to have a
choice \cite{houston} in selecting the collective phenomena  causing the
pseudogap, the comparison between theory and thermodynamic data provides a
consistency check, once the choice has been made. 

In this article we perform such a consistency check for a simple pair
fluctuation model of the pseudogap regime, introduced by Vilk and
Tremblay\cite{vilk}.  We opted for this formulation because 
it permits essentially analytical treatment, and therefore the
analysis is not obscured the numerical difficulties ubiquitous in most
other incarnations of the pair fluctuation scenario \cite{randeria}. We believe
however that the results we obtained have a range of relevance that goes beyond
the present formulation and it is characteristic of a larger class 
of pair fluctuation models for the pseudogap \cite{levin-cv}.

In our calculation we consider beside the usual single particle
contribution \cite{agd}, the contribution of the
fluctuating pairs. We show that (a) it has the same order of magnitude
as the single particle contribution and, (b) it is essential 
for explaining the {\em universal} presence of a broad hump in the
specific heat data.  In what follows we first present the main
theoretical framework, and then report a detailed analysis of the
experimental data within this framework. The central result of our 
analysis is the doping dependence of the {\em pairing} pseudogap
energy extracted from specific heat data, which compares well with the doping dependence of the
pseudogap scale inferred from a variety of other measurements.

\emph{Pairing contribution. }  For the calculation of the pairing contribution
we start with the usual expression of the
grand potential for pair fluctuations \cite{nozieres}
\begin{equation} 
\Omega _{p}\left(T\right) =-T\sum\limits_{{\mathbf{k}},\omega _n,\sigma 
}\ln \left[ 1-g\chi \left( {\mathbf{q}},\omega _n\right) \right],   \label{6} 
\end{equation} 
where $\chi({\mathbf{q}},\omega _n)$ is the pairing susceptibility. In what
follows we will consider only the non-interacting, bare susceptibility:

\begin{equation}
\chi({\mathbf{q}},\omega _n)
=\frac{1}{\beta}\sum_{{\mathbf{k}},\zeta_m}G^0({\mathbf{k}}+{\mathbf{q}},i\zeta_m+i\omega_n)
G^0({\mathbf{k}},i\zeta_m), 
\end{equation}
where $[G^0({\mathbf{k}},i\zeta_n)]^{-1}= i\zeta_n -\epsilon_{{\mathbf {k}}}$.
Here $\epsilon_{\mathbf{k}}$ is the single particle dispersion, $
\zeta_n=(2n+1)\pi/\beta$ is the fermionic Matsubara frequency. 
The summation in Eq. (\ref{6}) is done over
the pair  momentum $\mathbf{q}$, bosonic Matsubara frequencies $\omega_n=2\pi
n/\beta$ and spin $\sigma$.  We consider the renormalized classical regime
corresponding to  $\omega_n \sim 0$. Classical fluctuations give in
two-dimensions the dominant contribution to the self-energy at low
frequencies. Quantum fluctuations ($\omega_n \ne 0$) are important
only at low temperatures and are irrelevant for
temperatures $T_c <T<T^*$. Thus, in the renormalized classical regime the
susceptibility is given by   the relation:   
\begin{equation}
  1-g\chi \left( \mathbf{q}\right)
=e^{-\frac{2T^{*}}T}+\xi _0^2q^2.  \label{7}  
\end{equation}

The first term on the right side of the above expression corresponds
to an exponentially growing correlation length,  $\xi \sim \exp
(-T^*/T)$, a hallmark of the classical fluctuation regime. The onset of this
 regime happens at $T^{*}$, a characteristic
temperature scale set by the coupling strength $g$ and pseudogap $\Delta$
(note that  in this notations $g$ has units of energy):
\begin{equation}  
k_BT^* = \frac{2\pi}{g}\Delta^2.
\end{equation} 
$T^*$ can also be identified with the onset of pseudogap behavior. In
principle  both $\Delta$ and $g$ can be doping dependent. The dimensionless
specific heat coefficient $\gamma (T) \equiv c_v(T)/k_B^2TN(0)$ can be obtained
directly from Eq.(1)  according to the thermodynamic relation  
$c_v(T) = -T\partial ^2 \Omega /\partial T^2$. The result for the {\em pair
contribution} is 

\begin{equation}
\gamma_p(T)= 16\pi A\left(\frac{T^*}{T}
 \right)^3e^{-\frac{2T^*}{T}}\left(    1+\frac{T}{2T^*}\ln\Lambda   \right).
\label{8} \end{equation}

In strictly two dimensions the integration of the pair momenta needs to
be regularized by introducing an upper cutoff $\Lambda$. In practice, however, the materials we are
interested in have a large, but finite, c-axis anisotropy, that naturally cuts
off the pair momenta $\Lambda\sim \xi_0/d_c$ where
$d_c$ is the interlayer distance. In Eq. (\ref{8}) $A \sim
(\xi_0^2k_BT^*N(0))^{-1} $ is the parameter that  governs the relative
importance of the pairing contribution. The characteristic feature of the
pairing contribution to  the specific heat is  a broad hump near a temperature
$T\sim T^{*}$. This contribution in general, and the hump feature in
particular, are important for obtaining a  good fit to the experimental data.

\emph{Single particle contribution.} The single
particle contribution to the grand potential is given by\cite{agd}
\begin{equation} 
\Omega _{s}\left(T\right) =-T\sum\limits_{{\mathbf{k}},\omega 
_n,\sigma }\left\{ \ln \left[ G^{-1}\left( {\mathbf{k}},\omega _n\right) 
\right] -\Sigma \left( {\mathbf{k}},\omega _n\right) \cdot G\left(
{\mathbf{k}}  ,\omega _n\right) \right\}.   \label{1} 
\end{equation}
For the present purposes we identify the self-energy $\Sigma ({\bf k},
\omega_n)$ in the above expression with the pairing self energy
\cite{vilk} of the renormalized classical regime:
\begin{equation} 
\Sigma \left( \mathbf{k},\omega _n\right) =\frac{\Delta ^2}{i\omega _n+\epsilon _{\mathbf{k}}}.  \label{2} 
\end{equation} 
where $\Delta$ is the pseudogap energy scale. 
For the dimensionless specific heat coefficient coming from this contribution
we obtained the result: \begin{equation} 
\gamma_{s}(T) =\frac{2\pi^2}{3}+\int\limits_0^\infty\left(\frac{N\left(
x,d\right) } {N\left( 0\right)
}-1\right)f\left(x\right)\left(1-f\left(x\right)\right)x^3. \label{4} 
\end{equation} 

 In Eq.(\ref{4}) we introduced the following notations: 
$x=E_{\mathbf{k}}/k_BT$,  $d=\Delta /k_BT$,
$E_{\mathbf{k}}=\sqrt{\epsilon_{\mathbf{k}}^2+\Delta^2}$,   $f\left( x\right)
=\left( e^x+1\right) ^{-1}$.   As indicated by  angle-resolved photoemission
spectra \cite{arpes} taken in the pseudogap regime, the single particle states
in this regime are very broad.  During the analysis of the specific heat data
we found that even for near-optimal doping an intrinsic broadening of the
single particle states must be considered in order to obtain a reasonable
account of the data based on the single particle contribution
(\ref{4}).  The importance of broadening in the single
particle states becomes even more apparent when our analysis is performed on
data taken on underdoped samples. This trend is again consistent with that
observed in $ARPES$. We therefore introduce a phenomenological
broadening of the single particle density of states in the following simple
form \begin{equation} 
\frac{N\left( x,d\right) }{N\left( 0\right) }=\left| Re\frac{x-i\gamma }{% 
\sqrt{\left( x-i\gamma \right) ^2+d^2}}\right|   \label{5} 
\end{equation} 
and $\gamma=\Gamma /k_BT$ (not to be confused with $\gamma_p(T)$ or
$\gamma_s(T)$). We would like to note that while the broadening $\Gamma$ is
essential for treating the single particle contribution it is less important
for the qualitative behavior of the pair contribution $\gamma_p(T)$ as
function of temperature.  The total contribution to the specific heat
coefficient is  \begin{equation}
\gamma_{el}(T)=\gamma_{s}(T)+\gamma_{p}(T). \label{result} 
\end{equation} 

\emph{Relative magnitude of the pairing and single particle contributions. }
From the results obtained in Eqs. (4) and (7) we see that the relative
importance of pairing  contribution $\gamma_p (T)$ compared to the usual
contribution $\gamma_s (T)$ is governed by the parameter $A\sim
(\xi_0^2k_BT^*N(0))^{-1}$. Using $\xi_0\sim 10 A $, $T^*\sim 100K $
and typical density of states, (for example, from $ARPES$ data  we get
$3$  $states/eV$ per $Cu-O$ in a unit cell \cite{arpes}), we find that
$\gamma_p$ and $\gamma_s$ are comparable in magnitude: $\gamma_p (T)\sim
\gamma_s(T)$. In contrast, for a superconductor with long coherence length such
as $Al$ $(\xi_{Al}\sim 100\rm{\xi_{YBCO})}$ the pairing contribution $\gamma_p(T)$ is
entirely negligible.

\emph{Comparison with the experiment.} A typical comparison
of the experimental data with the theoretical result is presented in $Fig.1$. 
The single particle and the pairing contribution is shown
separately in order to emphasize the relative size and the importance of {\em
  both} terms.
\begin{figure}

  \centerline{\includegraphics[width=3.5in]{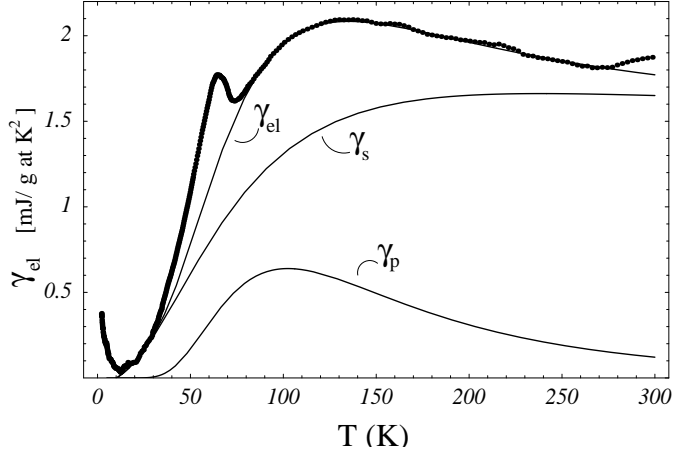}} 

\vspace*{3ex}

\caption{Fit to the experimentally measured specific heat
of $YB_{2}Cu_{3}O_{6.73}$ (dots) using  the theoretically  calculated  single
particle and pairing contributions (lines). Note that the single   particle
contribution $\gamma_s (T)$ cannot reproduce the hump and the pairing term
$\gamma_p (T)$ is necessary.} 

\label{fig:spt}

\end{figure}

The single particle contribution monotonically 
increases with increasing temperature and  by itself cannot explain the hump
observed in the experimental data. If, however, the pairing contribution
$\gamma_p (T)$ (with  maximum near $T^*$) is included, the fit is reasonable
\cite{noi}. During the fitting procedure only the normal state $( T>T_c)$
specific heat values are used. 
 \begin{figure}

  \centerline{\includegraphics[width=3.5in]{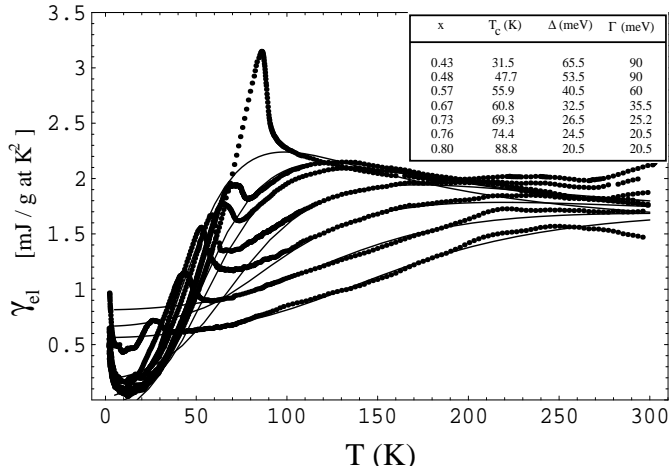}} 

\vspace*{3ex}

\caption{Specific heat data (dots) for
$YB_{2}Cu_{3}O_{6+x}$ for a wide range of doping (from below) $x=0.43$, $0.48$,
$0.57$, $0.67$,$0.73$,$0.76$ and $0.80$, and the corresponding theoretical
curves (lines). The inset gives extracted values for $\Delta$ and $\Gamma$. } 

\label{fig:ybco}

\end{figure}
An overall factor of order of $0.1$ and dimensions
of $ mJ/gatK^2 $ is used to scale the dimensionless theoretical expressions to
the data at optimal doping. Once this constant is obtained as described above,
[ $0.14$ $ mJ/g atK^2$ for $La_{2-x}Sr_xCuO_4$
and $0.23$ $ mJ/g atK^2$ for $YBa_2Cu_3O_{6+x}$ ] 
it is no longer a fitting parameter and remains fixed for all
doping levels.
When considering both single particle and pair contributions, a direct
comparison of the theoretical results and the experimental data can
be performed. The results are presented in $Fig. 2$ for
$YBa_2Cu_3O_{6+x}$, and in $Fig. 3$ for $La_{2-x}Sr_{x}CuO_{4}$ \cite{bsco}. 

\begin{figure}

  \centerline{\includegraphics[width=3.5in]{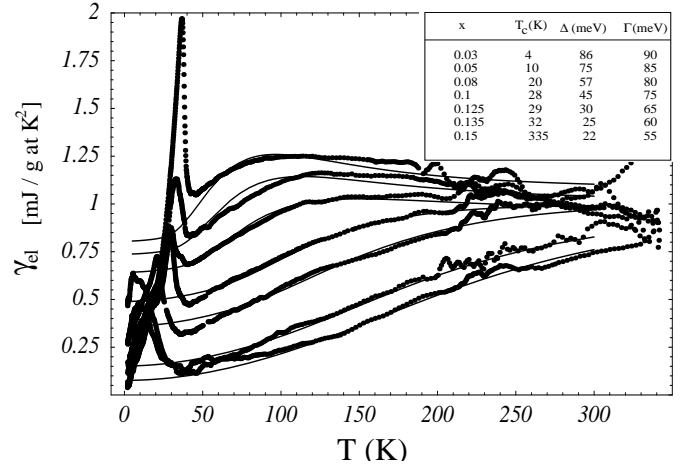}} 

\vspace*{3ex}

\caption{ Same as in Fig. 2, but now for $La_{2-x}Sr_{x}CuO_{4}$  [$x=0.03$,
$0.05$, $0.084$,   $0.1$, $0.125$, $0.135$ and $0.15$].} 

\label{fig:lsco}

\end{figure}

The phase diagram obtained from the extracted values of the pairing
pseudogap $\Delta$, and the experimental transition temperatures, is 
presented in $Fig.4$. Here we use reduced quantities (normalized to
their value at optimal doping) to present on the same scale the experimentally
measured  critical temperatures $T_c$ of $YBa_2Cu_3O_{6+x}$ and $ 
La_{2-x}Sr_xCuO_4$ compounds, and the corresponding pairing pseudogap scale $ 
\Delta$ extracted from data. 

\begin{figure}

  \centerline{\includegraphics[width=3.2in]{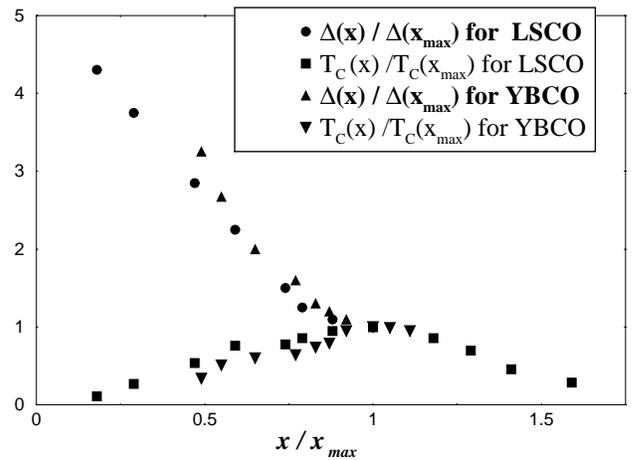}} 

\vspace*{3ex}

\caption{Doping dependence of the pairing pseudogap $\Delta$ extracted from
specific heat and of the  critical temperature $T_c$ for $YBa_2Cu_3O_{6+x}$
and $La_{2-x}Sr_xCuO_4$. }

\label{fig:phase}

\end{figure}

An interesting result emerges from the present
analysis.  The best fit is obtained if $T^*(x)$ is taken to be
{\em proportional} to $\Delta(x)$: $T^*\sim \Delta(x)$.

\begin{figure}

  \centerline{\includegraphics[width=3.2in]{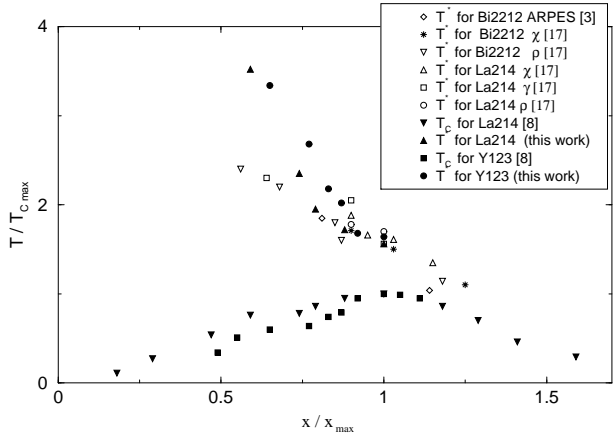}} 

\vspace*{3ex}

\caption{ $T^*$ and $T_c$ dopind dependence measured as $ARPES$, $\gamma$,
$\chi$ for $YBa_2Cu_3O_{6+x}$, $La_{2-x}Sr_xCuO_4$ and $
Bi_2Sr_2CaCu_2O_{8+x}$. }   

\label{fig:phase1}

\end{figure}

This proportionality has also been indicated in Ref. 
 \cite{oda} based on a scaling analysis of various experimental
 data, and an approximate ratio of $\Delta=1.5T^*$ has been found, similar to
our finding. Eq.(3) implies, at least for the present
pair fluctuation formalism, that the ratio $\Delta(x)/g(x)$ is independent of
doping. Thus the doping dependence of the $\Delta (x) /\Delta
(x_{max})$  coincides with  doping dependence of $T^*(x)/T^*(x_{max})$. The
broadening $\Gamma (x)$ increases with decreasing the doping, which is again in
qualitative agreement with the $ARPES$ results \cite{arpes}.  

\emph{Conclusions.} We presented the first systematic theoretical 
analysis of the electronic specific heat in the pseudogap state. While this 
analysis was performed within the classical pair 
fluctuation framework for the pseudogap, we suspect that the
general  features of the results are relevant for a large class of pairing
fluctuation scenarios. 

The main results of this analysis can be summarized as
follows: $(i)$ both the single particle and pair contributions are of the
same order  of magnitude and needed to fit the specific heat data of pseudogapped samples. $(ii)$ the electronic specific
heat data of underdoped cuprate samples have several universal features that
are well captured by our pair fluctuation model. Our analysis of the
experimental data yielded a pairing pseudogap energy scale $\Delta(x)$ that
has a magnitude and  doping dependence which is similar to the pseudogap energy
scales extracted from  other experiments (cf. $Fig.5$). Our main result
is the  phase diagram presented in $Fig.4$.
 
\emph{Acknowledgments.} This work benefited very much from Dr. John W. 
Loram's kind offer to make his specific heat data \cite{loram} available, and from several discussions on this issue with him and Dr. 
Jeffrey Tallon. We would also like to thank Prof. Alexei A. Abrikosov, Dr. 
Ioan Kosztin, Prof. Migaku Oda, Prof. John Zasadzinski, and especially Dr. 
Michael R. Norman for helpful discussions. This research was supported in 
part by the NSF under awards DMR91-2000 (administrated through the Science 
and Technology Center for Superconductivity), and the U.S. DOE, BES, under 
Contract No.~W-31-109-ENG-38.


\begin{thebibliography}{99}
\bibitem{noi} Note that in our framework $T_c=0$. We do not address the position and magnitude of the specific heat
anomaly near the experimentally observed $T_c$.
\bibitem{bsco} Since the very recent specific heat data
{\protect\onlinecite{houston}} taken on $Bi_2Sr_2CaCu_2O_{8+x}$ is similar with
the one in $Fig.2$ and $Fig.3$, we anticipate
that the analysis of $Bi_2Sr_2CaCu_2O_{8+x}$ data will yield qualitatively
identical result to ones obtained here for $YBa_2Cu_3O_{6+\delta}$ and
$La_{2-x}Sr_xCuO_4$.
\end{thebibliography}
\end{document}